\RequirePackage{fix-cm}
\documentclass[a4paper,11pt]{article}

\usepackage{authblk}
\usepackage{amsfonts,amsmath,amssymb,amsthm}
\usepackage{algorithm}
\usepackage{paralist}
\usepackage{booktabs}
\usepackage{url}
\usepackage{hyperref}
\usepackage{graphicx}
\usepackage{microtype}
\usepackage{url}
\usepackage{hyperref}
\usepackage{tabularx}
\usepackage{multirow}
\usepackage{subfig}
\usepackage{tikz}
\usetikzlibrary{positioning}
\usetikzlibrary{decorations.pathreplacing}

\newcommand{\N}{\mathbb{N}}
\newcommand{\Z}{\mathbb{Z}}
\newcommand{\F}{\mathbb{F}}

\newtheorem{lemma}{Lemma}
\newtheorem{theorem}{Theorem}

\newtheorem{example}{Example}

\tikzset{
    mybrace/.style={decorate,decoration={brace,aspect=#1}}
}

\providecommand{\keywords}[1]{\textbf{\textit{Keywords }} #1}

\begin{document}

\title{Bent Functions in the Partial Spread Class Generated by Linear Recurring Sequences}

\author[1]{Maximilien Gadouleau}
\author[2]{Luca Mariot}
\author[3]{Stjepan Picek}
	
\affil[1]{{\small Department of Computer Science, Durham University, South Road, Durham, DH1 3LE, UK} 
	
	{\small \texttt{m.r.gadouleau@durham.ac.uk}}}

\affil[2]{{\small Cyber Security Research Group, Delft University of Technology, Mekelweg 2, Delft, The Netherlands} 
	
	{\small \texttt{l.mariot@tudelft.nl}}}

\affil[3]{{\small Digital Security Group, Radboud University, PO Box 9010, 6500 GL, Nijmegen, The Netherlandss} 
	
	{\small \texttt{stjepan@computer.org}}}

\maketitle

\begin{abstract}
We present a construction of partial spread bent functions using subspaces generated by linear recurring sequences (LRS). We first show that the kernels of the linear mappings defined by two LRS have a trivial intersection if and only if their feedback polynomials are relatively prime. Then, we characterize the appropriate parameters for a family of pairwise coprime polynomials to generate a partial spread required for the support of a bent function, showing that such families exist if and only if the degrees of the underlying polynomials is either $1$ or $2$. We then count the resulting sets of polynomials and prove that for degree $1$, our LRS construction coincides with the Desarguesian partial spread. Finally, we perform a computer search of all $\mathcal{PS}^-$ and $\mathcal{PS}^+$ bent functions of $n=8$ variables generated by our construction and compute their 2-ranks. The results show that many of these functions defined by polynomials of degree $b=2$ are not EA-equivalent to any Maiorana-McFarland or Desarguesian partial spread function.
\end{abstract}


\section{Introduction}
\label{sec:intro}

Boolean functions play an important role in cryptography, coding theory, and combinatorial designs~\cite{mesnager16}. Among them, \emph{bent functions} are of particular interest since they lie at the highest possible Hamming distance from the set of all affine functions, or equivalently they reach the highest possible nonlinearity. For this reason, bent functions have been extensively used in the past for designing stream and block ciphers since highly nonlinear Boolean functions are useful to withstand fast-correlation and linear cryptanalysis attacks. Even though bent functions are unbalanced, highly nonlinear balanced functions can be derived from them~\cite{dobbertin94}. For this reason, bent functions have been used in the past for designing stream and block ciphers since highly nonlinear Boolean functions are useful to withstand fast-correlation and linear cryptanalysis attacks~\cite{carlet21}. Besides cryptography, bent functions are connected in coding theory to the \emph{covering radius} of first-order Reed-Muller codes, whose codewords are affine Boolean functions.

Over the last decades, many constructions of bent functions have been described in the related literature (see, e.g.,~\cite{carlet-bool,mesnager16,carlet21} for a survey of the main ones). A distinction is usually made between the \emph{primary} and \emph{secondary} constructions. Primary constructions build sets of bent functions from scratch, usually by leveraging on related combinatorial structures. Some of the most well-known primary constructions for bent functions include the \emph{Maiorana-McFarland construction}~\cite{mcfarland73}, which exploits permutations over $\F_2^{n}$, and \emph{Dillon's construction}~\cite{dillon74}, based on the class of \emph{partial spreads} $\mathcal{PS}$. On the contrary, \emph{secondary constructions} build new bent functions starting from existing ones. For example, the \emph{Rothaus's construction}~\cite{rothaus76} takes three bent functions of $n$ variables whose sum is also bent and yields a bent function of $n+2$ variables.

The search for novel methods to design bent functions is still an interesting and active research area nowadays, for a twofold motivation:
\begin{itemize}
	\item \emph{Discovering new functions}. Notwithstanding the multitude of existing constructions, they only cover a tiny fraction of the total number of bent functions~\cite{mesnager16}, and the complete enumeration of bent functions remains an open question for $n\ge 10$ variables~\cite{polujan20}. Therefore, finding new constructions that yield previously unknown bent functions is still an interesting research avenue to pursue. However, one must remark that this direction is becoming increasingly difficult precisely because many constructions are already in place. This makes the discovery of new bent functions both unlikely and cumbersome since, in principle, one has to check inequivalence against a large number of known classes.
	\item \emph{Finding new constructions for known functions}. Novel constructions that generate already known bent functions are an interesting research line as well, for several reasons. For example, from an implementation point of view, such constructions could highlight more efficient ways to design the corresponding bent functions, other than by classic lookup tables. More generally, a novel construction could provide a new perspective on understanding the structure of a known class of bent functions and spawning new research questions linked both to the construction of new bent functions and other interesting combinatorial objects. As we will argue in the following, we deem our work an example of this approach.
\end{itemize}

In this paper, we present a new primary construction of bent functions in the partial spread class $\mathcal{PS}$ by using the subspaces spanned by \emph{Linear Recurring Sequences} (LRS) over finite fields. The main idea is to define a linear mapping through the feedback polynomial of an LRS and then to use its kernel as a subspace in a partial spread. The main contributions of this paper can be summarized as follows:
\begin{itemize}
	\item We prove that the kernels of two linear mappings have a trivial intersection if and only if the feedback polynomials of their LRS are pairwise coprime.
	\item We show that a family of pairwise coprime polynomials large enough to define the partial spread for a bent function exists if and only if the degree of the involved polynomials is either $b=1$ or $b=2$, assuming that all polynomials have a nonzero constant term.
	\item We prove that for degree $b=1$, the functions given by our LRS construction coincide with the Desarguesian partial spread class.
	\item We perform a computer search of all bent functions of $n=6,8$ variables generated by our LRS construction, remarking that they always have maximal degree $n/2$, even for $\mathcal{PS}^+$-type functions.
	\item We analyze the distribution of the $2$-ranks for the LRS bent functions of $n=8$ variables. For the degree $b=1$, we independently verify the distribution reported by Weng et al.~\cite{weng07} for functions in the Desarguesian partial spread. For degree $b=2$, we remark that many of the obtained bent functions have a rank higher than 42, and thus they are not EA-equivalent to any Maiorana-McFarland or Desarguesian partial spread function.
\end{itemize}

The remainder of this work is structured as follows. Section~\ref{sec:background} reviews the background definitions on bent functions and linear recurring sequences used throughout the paper. Section~\ref{sec:construction} defines our LRS construction, proving that the kernels of two LRS linear mapping have a trivial intersection if and only if the associated feedback polynomials are coprime. Section~\ref{sec:count} characterizes the families of pairwise coprime polynomials that are required for the LRS construction and provides the corresponding counting result. Section~\ref{sec:eq-ds} shows that the LRS construction equals the Desarguesian partial spread construction when using polynomials of degree $1$. Section~\ref{sec:comp-res} discusses the computer search experiments for bent functions of $n=6,8$ variables generated by the LRS construction, reporting the distribution of the $2$-ranks. Finally, Section~\ref{sec:conclusions} summarizes the main results presented in this paper and points out several avenues for future research, discussing the connection with the cellular automata approach used in~\cite{gadouleau20}.

\section{Background}
\label{sec:background}

This section covers the necessary background notions used throughout the paper. We begin by introducing the basic definitions and results related to bent Boolean functions, describing the main known primary constructions (namely, the Maiorana-McFarland construction and Dillon's partial spread class), the extended affine equivalence relation, and a to check (in)equivalence of a bent function against a class of other known functions. We then move to linear recurring sequences and their vector spaces, representing the main combinatorial objects used to define our new construction of bent functions.

\subsection{Bent Functions}
\label{subsec:bent}

We refer the reader to~\cite{carlet21} for a thorough treatment of the results recalled in this section about Boolean functions. In what follows, let $\F_q$ be the finite field with $q$ elements (where $q=p^{\alpha}$ is a power of a prime number), and denote by $\F_q^n$ the $n$-dimensional vector space over $\F_q$, with $\underbar{0}$ being its null vector. For $q=2$, sum and multiplication on $\F_2$ correspond to the XOR and logical AND operations, respectively. Following the literature convention about Boolean functions, we will denote the sum operation over $\F_2$ by $\oplus$, while for a generic finite field $\F_q$ we will adopt the normal sum symbol $+$. 
On the other hand, we will denote the multiplication operation in all finite fields by concatenation of the operands. A \emph{Boolean function} of $n$ variables is a mapping $f: \F_2^n \to \F_2$. The most natural way to uniquely represent a Boolean function $f$ is by means of its \emph{truth table}, which is the vector $\Omega_f \in \F_2^{2^n}$ that lists the output of $f$ evaluated over all $2^n$ input vectors $x \in \F_2^n$ in lexicographic order. The \emph{support} of $f$ is the subset of input vectors that map to $1$, that is, $supp(f) = \{x \in \F_2^n : f(x) \neq 0\}$, while the \emph{Hamming weight} of $f$ is defined as $w_H(f) = |supp(f)|$, i.e., the number of ones in the truth table of $f$. Functions with the Hamming weight equal to $w_H = 2^{n-1}$ are also called balanced, since their truth table is composed of an equal number of zeros and ones, and they play an important role in the design of stream and block ciphers~\cite{carlet21}. The \emph{polarity truth table} $\Omega_{\hat{f}}$ of $f: \F_2^n \to \F_2$ is the truth table of the function $\hat{f}: \F_2^n \to \{-1,+1\}$ defined as $\hat{f}(x) = (-1)^{f(x)}$ for all $x \in \F_2^n$.

The \emph{Algebraic Normal Form} (ANF) is another useful representation that expresses a Boolean function $f: \F_2^n \to \F_2$ as a multivariate polynomial over the quotient ring $\mathbb{F}_2[x_1,\cdots,x_n]/(x_1^2 \oplus x_1, \cdots, x_n^2 \oplus x_n)$:
\begin{equation}
P_f(x) = \bigoplus_{I \in \mathcal{P}([n])} a_I \left( \prod_{i \in I} x_i \right) \enspace ,
\end{equation}
with $\mathcal{P}([n]) = 2^{[n]}$ being the power set of $[n] = \{1,\cdots,n\}$. The \emph{algebraic degree} of $f$ is defined as the cardinality of the largest subset $I$ such that $a_I \ne 0$. In particular, \emph{affine functions} are defined as those Boolean functions with degree at most $1$. Notice that the ANF is a unique representation of a Boolean function, and in particular one can retrieve the truth table back from the ANF coefficients through the \emph{M\"{o}bius transform}:
\begin{equation}
\label{eq:mobius}
f(x) = \bigoplus_{I \in \mathcal{P}[n]: I \subseteq supp(x) } a_I \enspace ,
\end{equation}
A third common method to uniquely represent Boolean functions used in cryptography is the \emph{Walsh-Hadamard transform}. Formally, the Walsh-Hadamard transform of a Boolean function $f:\F_2^n \to \F_2$ is the mapping $W_f: \F_2^n \to \Z$ defined for all $a \in \F_2^n$ as
\begin{equation}
\label{eq:walsh}
W_f(a) = \sum_{x \in \F_2^n} (-1)^{f(x) \oplus a \cdot x} \enspace ,
\end{equation}
where $a \cdot x  = \bigoplus_{i=1}^n a_ix_i$ is the \emph{scalar product} between $a$ and $x$. One may easily see that a function $f$ is balanced if and only if its Walsh-Hadamard transform vanishes on the null vector, i.e., if and only if $W_f(\underbar{0}) = 0$. In particular, the Walsh-Hadamard coefficient $W_f(a)$ quantifies the correlation between $f$ and the linear function $a\cdot x$. The lower the absolute value of $W_f(a)$, the lower will be the correlation of $f$ from $a\cdot x$ (and from its affine counterpart $1 \oplus a\cdot x$), and thus the higher will be the Hamming distance between the truth tables of the two functions. In particular, the \emph{nonlinearity} of a Boolean function $f:\F_2^n \to \F_2$ is defined as the minimum Hamming distance of $f$ from the set of all affine functions, and it can be computed as follows:
\begin{equation}
\label{eq:nl}
Nl_f = 2^{n-1} - \frac{1}{2}\max_{a \in \F_2^n}\left(|W_f(a)|\right) \enspace .
\end{equation}

Therefore, a Boolean function with high nonlinearity must be characterized by a low maximum absolute value among its Walsh-Hadamard coefficients. \emph{Parseval's relation} states that the sum of the squared Walsh-Hadamard spectrum is constant for any Boolean function $f: \F_2^n \to \F_2$, and it equals:
\begin{equation}
\label{eq:parseval}
\sum_{a \in \F_2^n} [W_f(a)]^2 = 2^{2n} \enspace .
\end{equation}

From Parseval's relation, one can remark that the lowest maximum absolute value of the Walsh-Hadamard transform occurs when the constant $2^{2n}$ is uniformly ``spread'' among all $2^n$ coefficients, that is when each coefficient equals $2^{\frac{n}{2}}$ in absolute value. This observation yields the \emph{covering radius bound} for the nonlinearity of an $n$-variable Boolean function:
\begin{equation}
\label{eq:cov-bound}
Nl_f \le 2^{n-1} - 2^{\frac{n}{2}-1} \enspace .
\end{equation}
Functions satisfying with equality Equation~\eqref{eq:cov-bound} -- or equivalently, whose Walsh-Hadamard coefficients all equal $2^{\frac{n}{2}}$ in absolute value -- are called \emph{bent functions}. When $n$ is even, such functions exist since the Walsh-Hadamard coefficients must be integer numbers. Although achieving the highest possible nonlinearity granted by the covering radius bound, bent functions cannot be employed directly in the design of stream or block ciphers since they are always imbalanced. As a matter of fact, we have $W_f(\underbar{0}) = \pm 2^{\frac{n}{2}}$ for any bent function, which means that its Hamming weight is $2^{n-1} \pm 2^{\frac{n}{2}-1}$.

There are several ways to construct bent functions proposed in the literature. Usually, such methods are divided in \emph{primary} and \emph{secondary constructions}. Recall, a primary construction builds ``from scratch'' new bent functions by leveraging other kinds of combinatorial objects. On the other hand, a secondary construction derives new bent functions starting from already existing ones. This paper focuses on the former case.

One of the main primary constructions investigated in in the literature is the \emph{Maiorana-McFarland construction}, which is the set of all bent functions $f: \F_2^n \to \F_2$, with $n=2m$, defined as: 
\begin{equation}
\label{eq:mm}
f(x,y) = x \cdot \pi(y) \oplus g(y) \enspace ,
\end{equation}
for all $x,y \in \F_2^{m}$, where $\pi: \F_2^{m} \to \F_2^{m}$ is any permutation of $\F_2^{m}$. Therefore, for any $m \in \N$ there are $(2^m)!$ bent functions of $2m$ variables in $\mathcal{M}$.

A second well-known primary construction that give rise to a large number of bent functions was introduced by Dillon in his PhD thesis~\cite{dillon74}, and it is based on \emph{partial spreads}. A partial spread of $\F_2^n$, with $n=2m$, is a family $P$ of $m$-dimensional subspaces $S_1,S_2,\cdots,S_t \subseteq \F_2^n$ with pairwise trivial intersection (i.e., for all $i\neq j$ one has $S_i \cap S_j = \{\underbar{0}\}$). Further, a partial spread is a \emph{spread} if the union of its subspaces results in the whole space $\F_2^n$. The main result proved by Dillon is that one can construct a bent function $f: \F_2^n \to \F_2$, with $n=2m$, from a partial spread $P$ of $\F_2^n$ by defining the support of $f$ as the \emph{union} of the subspaces in $P$. Remark that the partial spread must be large enough to reach the Hamming weight required for a bent function. In particular, a bent function $f: \F_2^n \to \F_2$, $n=2m$, belongs to the class $\mathcal{PS}^{-}$ if $f(0)=0$ and its support is the union of $t=2^{m-1}$ subspaces of a partial spread $P$ of $\F_2^n$. Functions in the $\mathcal{PS}^-$ reach the maximum possible algebraic degree for a bent function of $n=2m$ variables, namely $m$. Bent functions belonging to the class $\mathcal{PS}^{+}$ are defined similarly, with $f(0)=1$ and their support being the union of $t = 2^{m-1}+1$ $m$-dimensional subspaces of a partial spread of $\F_2^n$. Contrarily to $\mathcal{PS}^-$, functions in $\mathcal{PS}^+$ can have algebraic degrees other than $m$. The union of $\mathcal{PS}^-$ and $\mathcal{PS}^+$ gives the whole partial spread class $\mathcal{PS}$.

Currently, the structure of the class $\mathcal{PS}$ is still far from being completely characterized, and several methods have been investigated to define partial spreads that are large enough to obtain $\mathcal{PS}^-$ and $\mathcal{PS}^+$ bent functions. Here, we introduce only the \emph{Desarguesian spread}, which is perhaps the best-known example of spread used to construct $\mathcal{PS}^-$ bent functions (see, e.g.,~\cite{carlet16} for a general overview of other partial spreads). Given $n=2m$, one can use the \emph{bivariate form} to represent the Desarguesian spread~\cite{mesnager16}. The vector space $\F_2^n$ is identified with the Cartesian product $\F_{2^m} \times \F_{2^m}$, and the Desarguesian spread is defined as:
\begin{align}
\label{eq:des-biv}
\nonumber
DS &= \{E_a \subseteq \F_{2^m} \times \F_{2^m}: \ a \in \F_{2^m}\} \cup E_{\infty} \enspace \textrm{, where}:  \\
\nonumber
E_a &= \{(x,ax) \in \F_{2^m} \times \F_{2^m}: \ x \in \F_{2^m} \} \enspace , \\
E_{\infty} &= \{(0,y) \in \F_{2^m} \times \F_{2^m}: \ y \in \F_{2^m}\} \enspace .
\end{align}
Then, any subset of $2^{m-1}$ elements of $DS$ is a partial spread whose union defines the support of a bent function. More in particular, these functions belong to the so-called class $\mathcal{PS}_{ap}$ (where $ap$ stands for ``affine plane''), which is a subset of $\mathcal{PS}^-$. Besides reaching maximal degree $n/2$, functions in the $\mathcal{PS}_{ap}$ class have the additional interesting property of being \emph{hyper-bent}, as shown, e.g., in~\cite{carlet06}. A Boolean function $f: \F_{2^n} \to \F_2$, $n$ even, is called hyper-bent if the function $f(x^i)$ is bent for all exponent $i$ coprime with $2^n-1$~\cite{youssef01}. As such, hyper-bent functions have the highest possible distance not only from all affine functions (which corresponds to the case $i=1$), but also from all \emph{bijective monomial functions}.

Given the great variety of primary constructions available in the literature, a crucial question when investigating a new construction is to assess whether the bent functions produced by it are essentially different from those belonging to other known classes. This is accomplished by using equivalence relations. The underlying idea is to classify the bent functions produced by the known constructions up to equivalence and then verify if the bent functions generated by a new construction belong to any of these classes or to different ones. The main equivalence relation used in this context is the \emph{extended affine equivalence} (EA-equivalence). Two Boolean functions $f,g:\F_2^n \to \F_2$ are EA-equivalent if there exists a linear permutation $L: \F_2^n \to \F_2^n$, two vectors $u,v \in \F_2^n$, and an element $c \in \F_2$ such that:
\begin{equation}
\label{eq:ea-eq}
g(x) = f(L(x) \oplus u) \oplus (v\cdot x) \oplus c \enspace ,
\end{equation}
for all $x \in \F_2^n$. A bent function $f: \F_2^n \to \F_2$, with $n=2m$, belongs to the \emph{completed Maiorana-McFarland class} $\mathcal{M}^{\#}$ if it is EA-equivalent to a function in $\mathcal{M}$.

One possible method to check EA-equivalence of a function concerning other classes resorts to the notion of rank, introduced by Weng et al.~\cite{weng07}. More precisely, the 2-rank of a Boolean function $f:\F_2^n \to \F_2$ is the rank of the $2^n \times 2^n$ binary matrix $A_f$ whose rows and columns are indexed by the vectors of $\F_2^n$, and which is defined as $A_f(x,y) = f(x \oplus y)$ for all $x,y \in \F_2^n$. This matrix is also the incidence matrix of the symmetric 2-design associated with a Boolean function. Weng et al. proved that EA-equivalent bent functions have the same rank. Therefore, one can prove that two bent functions are not equivalent by checking that their ranks differ. The paper by Weng et al. further characterizes the lower and upper bounds for different types of bent functions. In particular, the rank of any Maiorana-McFarland bent function of $n=2m$ variables ranges between $LB_{\mathcal{M}} = 2m+2$ and $UB_{\mathcal{M}} = 2^{m+1}-2$. On the other hand, bent functions defined over the Desarguesian partial spread have ranks between $LB_{DS} = 2^{m+1}-2$ and $UB_{DS} = \sum_{i=0}^m \binom{m}{i}2^{\min\{i, m-i\}}$. An interesting consequence of the fact that the two intervals overlap only on $2^{m+1}-2$ is that almost all bent functions arising from the Desarguesian partial spread class are inequivalent to any Maiorana-McFarland function. Moreover, one can show that a bent function is inequivalent to all Maiorana McFarland and Desarguesian spread functions by showing that its rank is higher than $UB_{DS}$.

\subsection{Linear Recurring Sequences}
\label{subsec:lrs}

This section covers only the basic notions of linear recurring sequences essential to present our construction. An excellent overview of this topic can be found in the book by Lidl and Niederreiter on finite fields~\cite{lidl94}.

Let $a, a_0, \cdots, a_{d-1} \in \F_q$. A sequence $\{x_i\}_{i \in \N}$ of elements in $\F_q$ is called a \emph{linear recurring sequence} (LRS) of order $d \in \N$ if it satisfies the following relation: 
\begin{equation}
\label{eq:lrs}
a + a_0x_i + a_1x_{i+1} + ... + a_{d-1}x_{i+d-1} = x_{i+d} \enspace ,
\end{equation}
for all $i \in \N$. The first $d$ elements $x_0, \cdots, x_{d-1}$ act as the \emph{initial values} of the sequence, while all subsequent ones are determined by applying the linear recurrence defined in Equation~\eqref{eq:lrs}. In what follows, we will assume that $a=0$, i.e., that the LRS is \emph{homogeneous}. The \emph{feedback polynomial} of the LRS~\eqref{eq:lrs} is defined as:
\begin{equation}
\label{eq:fbpol}
f(X) = a_0 + a_1X + \cdots a_{d-1}X^{d-1}  \enspace ,
\end{equation}
that is, $f(X)$ is the polynomial in $\F_q[X]$ of degree $d-1$ whose monomials are defined by the coefficients of the LRS.

In this work, we are interested in a particular variant of LRS of order $d$: we start from a vector of initial values $x = (x_0,\cdots,x_{2(d-1)}) \in \F_q^{2(d-1)}$ and we apply Equation~\eqref{eq:lrs} on $x$ for $d-1$ times. In other words, we define the map $F: \F_q^{2(d-1)} \to \F_q^{d-1}$ as:
\begin{equation}
\label{eq:rule}
F(x_0,\cdots,x_{2(d-1)})_i = a_0x_i + a_1x_{i+1} + ... + a_{d-1}x_{i+d-1} \enspace ,
\end{equation}
for all output coordinates $i \in \{0,\cdots,d-1\}$. Since the map $F$ is linear, we can describe it as $F(x) = M_F \cdot x^\top$, where $M_F$ is a $(d-1) \times 2(d-1)$ matrix of the following form:
\begin{equation}
\label{eq:ca-matr-f}
M_F = 
\begin{pmatrix}
a_0    & \cdots & a_{d-1} & 0 & \cdots & \cdots & \cdots & \cdots & 0 \\
0      & a_0    & \cdots  & a_{d-1} & 0 & \cdots & \cdots & \cdots & 0 \\
\vdots & \vdots & \vdots & \ddots  & \vdots & \vdots & \vdots & \ddots & \vdots \\
0 & \cdots & \cdots & \cdots & \cdots & 0 & a_0 & \cdots & a_{d-1} \\
\end{pmatrix} \enspace ,
\end{equation}
Therefore, we can compactly represent the linear map $F$ by the coefficients of the feedback polynomial $f$, i.e., $f \mapsto M_F$. Notice that $M_F$ has the form of the \emph{parity-check matrix} of a cyclic code, with $f$ playing the role of the \emph{parity check polynomial}. However, the code associated with $f$ is not cyclic in general. This happens, in particular, if and only if the \emph{generator polynomial} (which is defined as the reciprocal of $f$) divides $X^N-1$, where $N=2(d-1)-1$. On the other hand, evaluating $F$ on a particular vector $x \in \F_q^{2(d-1)}$ corresponds to computing the \emph{syndrome} of $x$. In particular $x$ will be a valid codeword if and only if $M_F \cdot x^\top = \underline{0}$, i.e., if and only if $x$ belongs to the \emph{kernel} of $F$. In what follows, we will consider the kernels of the $\F_q$-subspaces associated with this kind of linear mappings to generate partial spreads. As a final note, remark that $F$ may also be regarded as a linear \emph{cellular automaton} (CA)~\cite{mariot17}. The connection between the LRS used to define bent functions in this paper, and CA will be briefly discussed in the conclusions.

\section{The LRS Construction}
\label{sec:construction}

The first step of our construction requires to characterize when the kernels of two LRS subspaces have a trivial intersection. The next result shows that this is equivalent to computing the greatest common divisor of the respective feedback polynomials. 

\begin{lemma}
	\label{lm:trv-int}
	Let $f,g \in \F_q[X]$ be two polynomials over $\F_q$ of degree $d$, respectively defined as:
	\begin{align}
	\label{eq:fg-poly}
	f(X) &= a_0 + a_1X + \cdots + a_{d-1}X^{d-1} \enspace , \\
	g(X) &= b_0 + b_1X + \cdots + b_{d-1}X^{d-1} \enspace ,
	\end{align}
	with $a_i,b_i \in \F_q$. Further, let $F,G: \F_q^{2(d-1)} \to \F_q^{d-1}$ be the linear maps defined by the polynomials $f$ and $g$, respectively. Then, the kernels of $F$ and $G$ have trivial intersection if and only if $\gcd(f,g) = 1$, i.e., if and only if $f$ and $g$ are relatively prime.
	
	\begin{proof}
		The linear maps $F$ and $G$ are respectively defined as $F(x) = M_F\cdot x^\top$ and $M_G \cdot x^\top$ for all $x \in \F_q^{2(d-1)}$, where $M_F$ and $M_G$ are the two $(d-1)\times 2(d-1)$ matrices of the form. Define now the linear function $H: \F_q^{2(d-1)} \to \F_q^{2(d-1)}$ as $H = M_H \cdot x^\top$ for all $x \in \F_q^{2(d-1)}$, where
		\begin{equation}
		\label{eq:ca-matr-h}
		M_H = \begin{pmatrix} M_F \\ M_G \end{pmatrix} = 
		\begin{pmatrix}
		a_0    & \cdots & a_{d-1} & 0 & \cdots & \cdots & \cdots & \cdots & 0 \\
		0      & a_0    & \cdots  & a_{d-1} & 0 & \cdots & \cdots & \cdots & 0 \\
		\vdots & \vdots & \vdots & \ddots  & \vdots & \vdots & \vdots & \ddots & \vdots \\
		0 & \cdots & \cdots & \cdots & \cdots & 0 & a_0 & \cdots & a_{d-1} \\
		b_0    & \cdots & b_{d-1} & 0 & \cdots & \cdots & \cdots & \cdots & 0 \\
		0      & b_0    & \cdots  & b_{d-1} & 0 & \cdots & \cdots & \cdots & 0 \\
		\vdots & \vdots & \vdots & \ddots  & \vdots & \vdots & \vdots & \ddots & \vdots \\
		0 & \cdots & \cdots & \cdots & \cdots & 0 & b_0 & \cdots & b_{d-1} \\
		\end{pmatrix} \enspace .
		\end{equation}
		In other words, the matrix $M_H$ is simply the \emph{superposition} of the two matrices $M_F$ and $M_G$. Remark that $M_H$ is also the \emph{Sylvester matrix} of the polynomials $f$ and $g$. It is a well-known fact that the determinant of the Sylvester matrix (also called the \emph{resultant} in this context) is nonzero if and only if $f$ and $g$ are relatively prime~\cite{gelfand08}. 
		
		Suppose now that $ker(F) \cup ker(G) = \{\underline{0}\} \subset \F_q^{d-1}$. Then, it follows that $ker(H) = \{\underline{0}\} \subset \F_q^{2(d-1)}$ as well. Indeed, assume that $ker(H)$ contains a nonzero vector $x' \in \F_q^{2(d-1)} \setminus \{\underline{0}\}$. Since $M_H$ is the superposition of $M_F$ and $M_G$, this implies that $x' \in ker(F)$ and $x' \in ker(G)$, contradicting the hypothesis that the two kernels have trivial intersection. Consequently, the kernel of $H$ has dimension zero, and by the rank-nullity theorem the dimension of the image of $H$ is $2(d-1)$, or equivalently $M_H$ is invertible and its determinant is nonzero. Thus, since $M_H$ is the Sylvester matrix of $f$ and $g$, it follows that $\gcd(f,g) = 1$.
		
		The other direction of the implication follows a symmetric reasoning: if $\gcd(f,g) = 1$ it follows that $M_H$ is invertible, and thus the kernel of $H$ is trivial. Hence, the intersection of the kernels of $F$ and $G$ is also the trivial space $\{\underline{0}\} \subset \F_q^{d-1}$ since $M_H$ is the superposition of $M_F$ and $M_G$. \qed
	\end{proof}
\end{lemma}

Therefore, we need to find a family of pairwise coprime polynomials of degree $b = d-1$ that is large enough to define a bent function. Following what we recalled in Section~\ref{subsec:bent}, for a $\mathcal{PS}^-$ function we need $t=2^{m-1}$ coprime polynomials of degree $b$. To this aim, let us take the finite field $\F_q$ with $q=2^l$, for $l \in \N$. This is because a partial spread for a bent function must be defined over the vector space $\F_2^n$, $n=2m$. In particular, each vector $x \in \F_{2^l}^{2b}$ must also be converted in a corresponding binary vector $x \in \F_2^n$ since the union of the vectors in the partial spread will form the support of the bent function. In other words, we require that $lb = m$. By identifying $\F_{2^l}$ with the vector space $\F_2^l$, a vector $x$ in $\F_{2^l}^{2b}$ is a $2b$-tuple whose components are in turn binary $l$-tuples:
\begin{equation}
\label{eq:vect-l-m}
x = ((x_{0,0}, \cdots, x_{0,l-1}), \cdots, (x_{2b-1,0}, \cdots, x_{2b-1,l-1})) \enspace .
\end{equation}
We now associate to each element $x \in \F_{2^l}^{2b}$ an element of $\F_2^{2lb}$ through the \emph{flattening} operator $\varphi: \F_{2^l}^{2b} \to \F_2^{2lb}$, which simply drops the parentheses inside the vector representation of $x$:
\begin{equation}
\label{eq:vect-lm}
\varphi(x) = (x_{0,0}, \cdots, x_{0,l-1}, \cdots, x_{2b-1,0}, \cdots, x_{2b-1,l-1}) \enspace .
\end{equation}
It is then easy to see that $\varphi$ is bijective. We can now characterize the partial spreads arising from our construction:
\begin{theorem}
	\label{thm:ps-lrs}
	Let $m, l, b \in \N$ such that $m = lb$. If there are $t=2^{lb-1}$ (respectively, $t=2^{lb-1}+1)$) coprime polynomials of degree $b$ over $\F_q$ where $q=2^l$, then there exists a partial spread $P$ over $\F_2^n$, $n=2m$, whose union of its subspaces defines a bent function in the class $\mathcal{PS}^-$ (respectively, $\mathcal{PS}^+$).
	\begin{proof}
		Let $f_1,\cdots, f_t$ be the coprime polynomials of degree $b$ over $\F_q$, and let $F_1,\cdots, F_t: \F_q^{2b} \to \F_q^b$ be the corresponding linear maps associated to them. Define the following family of subspaces of $\F_2^n$, with $n=2m=2lb$:
		\begin{equation}
		P = \{ \Phi(ker(F_i)) \subseteq \F_2^n: 1 \le i \le t\} \enspace ,
		\end{equation}
		where $\Phi(ker(F_i)) = \{y \in \F_2^n: y = \varphi(x), x \in ker(F_i)\}$, for $1 le i \le t$. In other terms, the subspace $\Phi(ker(F_i))$ is obtained by taking the kernel of $F_i$ and applying the flattening operator to each vector in it. Since the polynomials $f_1,\cdots, f_t$ are pairwise coprime, by Lemma~\ref{lm:trv-int}, the kernels of the $F_i$ have pairwise trivial intersection. Clearly, the same property holds for the subspaces $\Phi(ker(F_i))$ in $P$ since they are just a different representation of the same kernels through the flattening operator. Therefore, $P$ is a partial spread over $\F_2^n$, and depending on its size ($t=2^{lb-1}$ or $t=2^{lb-1}+1$), it can be used to define the support of a $\mathcal{PS}^-$ or $\mathcal{PS}^+$ bent function, respectively. \qed
	\end{proof}
\end{theorem}

In the remainder of this section, we show two examples of bent functions obtained through our construction.

\begin{example}
	\label{ex:bent-4}
	Let $m=2$, $n=2m = 4 $, $l=1$, and $b=2$. Since $lb=m$, in this case we need to find $t= 2^{m-1} = 2$ relatively prime polynomials $f_1,f_2 \in \F_2[X]$ of degree $b=2$ to apply our construction. Let $f_1(X) = X^2 + 1$ and $f_2(X) = X^2 + X + 1$. In this case, there is no need to apply the flattening operator, since the ground field for the polynomials is already $\F_2$. The two linear maps $F_1,F_2: \F_2^4 \to \F_2$ are respectively defined by the following two matrices:
	\begin{displaymath}
	\label{eq:mat90-150}
	M_{F_1} = 
	\begin{pmatrix}
	1 & 0 & 1 & 0 \\
	0 & 1 & 0 & 1
	\end{pmatrix}
	, \enspace M_{F_2} =
	\begin{pmatrix}
	1 & 1 & 1 & 0 \\
	0 & 1 & 1 & 1
	\end{pmatrix}
	\end{displaymath}
	The kernels of $F_1$ and $F_2$ are the following ones:
	\begin{align*}
	ker(F_1) &= \{0000, 1010, 0101, 1111 \}, \\
	ker(F_2) &= \{0000, 1011, 0110, 1101 \},
	\end{align*}
	which clearly have trivial intersection. Therefore, the union of $ker(F_1)$ and $ker(F_2)$ (excluding the null vector) define the support of Boolean function $g: \F_2^4 \to \F_2$ with the following truth table:
	\begin{displaymath}
	\Omega_g = (0, 0, 0, 0, 0, 1, 1, 0, 0, 0, 1, 1, 0, 1, 0, 1) \enspace .
	\end{displaymath}
	The ANF of $g$ is defined as follows:
	\begin{displaymath}
	g(x_1,x_2,x_3,x_4) = x_1x_3 \oplus x_2x_3 \oplus x_2x_4.
	\end{displaymath}
	It is possible to verify that this function is bent in a number of ways. For example, one can observe that $g$ is equivalent to the function $g'(x_1,x_2,x_3,x_4) = x_1x_2 \oplus x_2x_3 \oplus x_3x_4$, up to a permutation of the input variables (in particular, it suffices to permute $x_2$ with $x_3$). It is well known in the literature (see, e.g.,~\cite{stinson04}) that the function $g'(x_1,\cdots, x_n) = x_1x_2 \oplus x_2x_3 \oplus \cdots \oplus x_{n-1}x_n$ is bent for any $n \in \N$ even.
\end{example}

\begin{example}
	\label{ex:bent-4+}
	The bent function $g$ defined in Example~\ref{ex:bent-4} belongs to the $\mathcal{PS}^-$ class, since its support is the union of $2^{2-1} = 2$ subspaces of dimension 2 with trivial intersection. If we want to obtain a $\mathcal{PS}^+$ function, we need an additional polynomial of degree $b=2$ that is coprime both to $f_1$ and $f_2$. To this end, we can select for instance $f_3(X) = X^2$. The kernel of the associated linear map $F_3$ is as follows:
	\begin{align*}
	ker(F_3) &= \{0000, 0001, 0010, 0011 \} \enspace ,
	\end{align*}
	which again has trivial intersection with both $ker(F_1)$ and $ker(F_2)$. Therefore, we can define a $\mathcal{PS}^+$ bent function $h: \F_2^4 \to \F_2$ by setting $h(0000) = 1$ and defining the rest of its support as the union of the three kernels minus their trivial intersection. We thus obtain the following truth table:
	\begin{displaymath}
	\Omega_h = (1, 1, 1, 1, 0, 1, 1, 0, 0, 0, 1, 1, 0, 1, 0, 1) \enspace ,
	\end{displaymath}
	with the ANF of $h$ being:
	\begin{displaymath}
	h(x_1,x_2,x_3,x_4) = x_1x_2 \oplus x_1x_3 \oplus x_2x_3 \oplus x_2x_4 \oplus x_1 \oplus x_2 \oplus 1 \enspace .
	\end{displaymath}
	
\end{example}

\section{Counting Bent Functions in the LRS Construction}
\label{sec:count}

The first research question spawning from Theorem~\ref{thm:ps-lrs} is whether for all even $n \in \N$ there are at least $t=2^{m-1}$ (respectively, $t=2^{m-1}+1$) pairwise coprime polynomials of degree $b=m/l$ over $\F_{2^{l}}$ to construct a $\mathcal{PS}^-$ (respectively, $\mathcal{PS}^+$) bent functions. In what follows, we focus on the case of monic polynomials with \emph{nonzero constant term} that are pairwise coprime to exploit the counting results proved in~\cite{mariot20}. The authors proposed a construction for such families of polynomials based on the multiplication of two irreducible polynomials of degree $k$ and $b-k$, respectively. In particular, they showed that the maximum size of the families that can be generated through this construction equals:
\begin{equation}
\label{eq:char-max-mols}
N_b = I_b + \sum_{k=1}^{\lfloor \frac{b}{2} \rfloor} I_k \enspace .
\end{equation}
In the formula above, $I_k$ denotes the number of irreducible monic polynomials of degree $n$ and with nonzero constant term over $\F_q$, which is $I_k=q-1$ for $k=1$, while for $k\ge 2$ it is given by \emph{Gauss's formula}:
\begin{equation}
\label{eq:gauss}
I_{k} = \frac{1}{k} \sum_{d|k} \mu(d)\cdot
q^{\frac{k}{d}} \enspace ,
\end{equation}
with $\mu$ denoting the \emph{M\"{o}bius function}. Further, in~\cite{mariot20} it is proved that such construction is optimal, meaning that $N_b$ actually corresponds to the maximum size attainable by any family of monic coprime polynomials of degree $b$ with nonzero constant term over $\F_q$. Thus, one can study Equation~\eqref{eq:char-max-mols} with respect to the parameters $l$, $b$, and $m$ to address the existence question for families of polynomials that satisfy the conditions of Theorem~\ref{thm:ps-lrs}. We now characterize such families for the case of $\mathcal{PS}^-$ functions in terms of the degrees of their polynomials:
\begin{theorem}
	\label{thm:char-pol}
	Let $l,b,m \in \N$ such that $lb = m$, and let $q=2^l$. Then there exists a family of $t=2^{m-1}$ pairwise coprime polynomials of degree $b$ and nonzero constant term over $\F_q$ if and only if $b \in \{1,2\}$.
	
	\begin{proof}
		We need to show that $N_b \ge \frac{1}{2}q^b$ if and only if $b \le 2$. We first settle the cases of $b \le 4$ one by one.
		
		For $b=1$, we obtain 
		\[
		N_1 = I_1 = q - 1 \ge \frac{1}{2}q.
		\]
		For $b=2$, we obtain
		\[
		N_2 = I_2 + I_1 = \frac{1}{2}( q^2 - q ) + (q-1) = \frac{1}{2}q^2( 1 + q^{-1} - 2q^{-2} ) \ge \frac{1}{2}q^2.
		\]
		For $b = 3$, we obtain 
		\begin{align*}
		N_3 &= I_3 + I_1 = \frac{1}{3} \left( q^3 - q \right) + ( q - 1 )\\
		&< \frac{1}{3} q^3 \left( 1 + 2 q^{-2}  \right) \le \frac{1}{3} q^3 \frac{3}{2}\\
		&= \frac{1}{2} q^3.
		\end{align*}
		For $b = 4$, we obtain 
		\begin{align*}
		N_4 &= I_4 + I_2 + I_1 = \frac{1}{4} \left( q^4 - q^2 \right) + \frac{1}{2}(q^2 - q) + ( q - 1 )\\
		&< \frac{1}{4} q^4 \left( 1 + q^{-2} + 2q^{-3}  \right) \le \frac{1}{4} q^4 \frac{3}{2}\\
		&= \frac{3}{8} q^4.
		\end{align*}
		
		We now move on to the case where $b \ge 5$. Denoting the smallest nontrivial divisor of $b$ by $p$, we first get the following upper bound on $I_b$:
		\[
		I_b \le \frac{1}{b} \left\{ q^b - q^{b/p} + (q^{b/p - 1} + \cdots q + 1) \right\} < \frac{1}{b} q^b.
		\]
		We also obtain the following upper bound:
		\[
		\sum_{k=1}^{\lfloor b/2 \rfloor} I_k \le q^{\lfloor b/2 \rfloor + 1} \le q^{b-2} \le \frac{1}{4} q^b \enspace .
		\]
		Combining, we obtain
		\[
		N_b = I_b + \sum_{k=1}^{\lfloor b/2 \rfloor} I_k < q^b \left(\frac{1}{b} + \frac{1}{4} \right) < \frac{1}{2} q^b \enspace .
		\]
		\qed
	\end{proof}
\end{theorem}
Hence, bent functions can be obtained from our LRS construction for all number of variables $n = 2m$, where $m=l$ when $b=1$, and $m=2l$ when $b=2$. This leads us to the following counting result:
\begin{theorem}
	\label{thm:count-pol}
	Let  $l, m \in \N$ and $b \in \{1,2\}$ such that $lb = m$, and let $q=2^l$. Then, the number of $PS^{-}$ bent functions of $n=2m$ variables that can be obtained by Theorem~\ref{thm:ps-lrs} with polynomials of degree $b$ and nonzero constant term is $\binom{2^{m}-1}{2^{m-1}}$ when $b=1$ and
	\begin{equation}
	\label{eq:count-b-2}
	\sum_{A = 0}^{I_2} \binom{I_2}{A} \sum_{B=0}^{2^{m-1} - A} \binom{I_1}{B} \binom{I_1 - B}{2 (2^{m-1} - B - A)} \frac{ (2(2^{m-1} - B - A))! }{ (2^{m-1} - B - A)! 2^{2^{m-1} - B - A} },
	\end{equation}
	where $I_2 = \frac{1}{2}(q^2 - q)$ and $I_1 = q - 1$, when $b=2$.
	
	\begin{proof}
		By Theorem~\ref{thm:char-pol} $b=1$ and $b=2$ are the only cases we need to address. Let $b=1$ (and thus $m=l$). Then, by Equation~\eqref{eq:char-max-mols}, the largest family $\mathcal{F}_1$ of coprime polynomials of degree 1 with nonzero constant term over $\F_q$ is composed of $N_1 = q-1 = 2^{m}-1$ elements. The number of subsets of $2^{m-1}$ elements of $\mathcal{F}_1$ that can be selected to apply Theorem~\ref{thm:ps-lrs} is $\binom{2^{m}-1}{2^{m-1}}$. For $b=2$, any family of $t = 2^{m-1}$ coprime polynomials of degree 2 with nonzero constant term over $\F_q$ consists of:
		\begin{enumerate}
			\item $A \le I_2$ irreducible polynomials of degree $2$;
			
			\item $B \le I_1$ polynomials of the form $f^2$, where $f$ is an irreducible polynomial of degree $1$;
			
			\item $C = t - B - A$ polynomials of the form $gh$, where $g$ and $h$ are irreducible polynomials of degree $1$;
		\end{enumerate}
		and obviously, the same irreducible polynomial of degree $1$ only appears once. There are $\binom{I_2}{A}$ choices for the first part of the family, $\binom{I_1}{B}$ choices for the second part of the family, and
		\begin{displaymath}
		\frac{1}{ C! } \binom{I_1 - B}{ 2 } \binom{I_1 - B - 2}{ 2 } \dots \binom{I_1 - B - 2C + 2}{ 2 } = \binom{I_1 - B}{ 2C } \frac{ (2C)! }{ C! 2^C }
		\end{displaymath}
		choices for the third part of the family. Combining all three parts, we obtain the formula.
		\qed
	\end{proof}
\end{theorem}

The results above refer to the number of families of coprime polynomials with a nonzero constant term that is large enough to construct $\mathcal{PS}^-$ bent functions. Although such functions will be the focus of our computer investigations in the next sections, one could also augment such families with other types of polynomials, as long as they are pairwise coprime with all the others. This could be used, for instance, to construct further $\mathcal{PS}^-$ functions or $\mathcal{PS}^+$ functions with polynomials of degree $b=1,2$. Additionally, one could combine these other types of coprime polynomials with families of degrees higher than $2$.

One simple idea to achieve this is to augment each family with the constant polynomial $1$ and the polynomial $X^b$. Although the former is not of degree $b$ while the latter does not have a constant term, it is easy to see that they are coprime both among themselves and to all other polynomials in the families considered in Theorems~\ref{thm:char-pol} and~\ref{thm:count-pol}. This idea spawns from the orthogonal array (OA) characterization of our construction adopted in~\cite{gadouleau20}, where the first two columns of the OA corresponds to the LRS subspaces defined by $1$ and $X^b$. We will elaborate further on this connection in the conclusions section.

We already used in Example~\ref{ex:bent-4} the polynomial $X^2$ to construct a $\mathcal{PS}^+$ function of $4$ variables, by adding it to the family $\{X^2+1, X^2 + X + 1\}$. One could also add the constant polynomial $1$, thereby obtaining a family of $4$ coprime polynomials of degree $b=2$. Since to define a $\mathcal{PS}^+$ function of $4$ variables with our LRS construction we need $2^{2-1}+1 = 3$ pairwise coprime polynomials, we can build $\binom{4}{3} = 4$ $\mathcal{PS}^+$ functions by selecting all subsets of three polynomials in $\{1, X^2, X^2+1, X^2+X+1\}$. Alternatively, one could build $\binom{4}{2} = 6$ $\mathcal{PS}^-$ functions from this family, since in this case we only need a subset of two polynomials.

The next example shows how the two polynomials $1$ and $X^b$ can be used to augment a family of coprime polynomials with a nonzero constant term of degree $b>2$, so that we have enough of them to apply our construction.

\begin{example}
	\label{ex:bent-6}
	Let $m=3$, and $l,b$ such that $lb=m$. There are only two possibilities, namely $l=3$ and $b=1$, and $l=1$ and $b=3$. The first one is already covered by Theorem~\ref{thm:char-pol} since $b=1$. Let us consider the case $l=1$ and $b=3$. From Equation~\eqref{eq:char-max-mols}, we have $N_3 = I_3 + I_1 = 2 + 1 = 3$ coprime polynomials of degree $3$ over $\F_2$ with nonzero constant term, which are the following ones:
	\begin{align*}
	f_1(X) &= X^3 + X^2 + 1 \\
	f_2(X) &= X^3 + X + 1 \\
	f_3(X) &= (X+1)(X^2+X+1) = X^3 + X^2 + X + 1 \enspace .
	\end{align*}
	To obtain a $\mathcal{PS}^-$ (respectively, a $\mathcal{PS}^+$) function we need $2^{3-1} = 4$ (respectively, $2^{3-1}+1 = 5$) coprime polynomials. By adding $1$ and $X^3$ to the set $\{f_1,f_2,f_3\}$, we can thus build $\binom{5}{4} = 5$ $\mathcal{PS}^{-}$ functions and one $\mathcal{PS}^+$ function.
\end{example}

\section{Equivalence to $DS$ Functions for Degree $b=1$}
\label{sec:eq-ds}

We now show that our LRS construction coincides with the Desarguesian partial spread class when considering polynomials of degree $b=1$. In this case, to generate a bent function $f: \F_2^n \to \F_2$ of $n = 2m$ variables, by Theorem~\ref{thm:ps-lrs} we need to find a set of $t = 2^{m-1}$ irreducible polynomials of degree $1$ over $\F_{2^m}$. This basically amounts to choose a subset of cardinality $t$ from the family:
\begin{equation}
\label{eq:irr-1}
\mathcal{I}_1 = \{a + X \in \F_{2^m}[X]: a \in \F_{2^m}^* \} \enspace .
\end{equation}
Thus, let $P = \{f_1(X), \cdots, f_t(X)\}$ be a subset of $\mathcal{I}_1$. Recall that each polynomial is used as an abstract representation for the coefficients of a LRS of order $d=b+1$, used to define the corresponding linear map. In particular, for $f_i(X) = a_i + X$, we have that $F_i$ equals:
\begin{equation}
\label{eq:irr-1-loc}
F_i(x_1,x_2) = a_ix_1 + x_2 \enspace ,
\end{equation}
for all pairs $(x_1,x_2) \in \F_{2^m} \times \F_{2^m}$. By Theorem~\ref{thm:ps-lrs}, the kernels of $F_i \equiv f_i$ for $i \in \{1,\cdots,t\}$ form a partial spread, and each of them is obtained by taking all pairs $(x_1,x_2) \in \F_{2^m} \times \F_{2^m}$ such that $x_2 = a_ix_1$, since $\F_{2^m}$ is a field of characteristic $2$. We have that:
\begin{align}
\label{eq:ker-irr-1}
\nonumber
ker(F_i) &= \{(x_1,x_2) \in \F_{2^m} \times \F_{2^m}: x_2 = a_ix_1\} \\
&= \{(x, a_ix) \in \F_{2^m} \times \F_{2^m}: x \in \F_{2^m}\} = E_{a_i} \enspace ,
\end{align}
where $E_{a_i}$ is a member of the Desarguesian spread as defined by Equation~\eqref{eq:des-biv} in bivariate form. We have thus obtained the following result:
\begin{lemma}
	\label{lm:psap-irr-1}
	Let $f: \F_2^n \to \F_2$, $n=2m$, be a bent function defined as in Theorem~\ref{thm:ps-lrs} with degree $b=1$. Then, $f \in \mathcal{PS}_{ap}$.
\end{lemma}
Therefore, when considering the family $\mathcal{I}_1$ of $2^{l}-1$ irreducible polynomials of degree $1$ over $\F_{2^l}$ with the nonzero constant term, our LRS construction is a particular case of the partial spread induced by the Desarguesian spread. Further, the two classes coincide if one adds the polynomial $X$ to the family $\mathcal{I}_1$ since in that case, one can construct $\binom{2^l}{2^{l-1}}$ $\mathcal{PS}_{ap}$ functions.

However, for degree $2$, the above reasoning on the Desarguesian spread does not hold. When $b=2$, the LRS is defined by three coefficients instead of two, with the input vector of the linear map consisting of $4$ coordinates. Consequently, the LRS is evaluated over three variables $x_1,x_2,x_3$, and there does not seem to be a straightforward way to express the kernel of the linear map as a set of pairs of the type $(x,ax)$. To the best of our knowledge, there are no other constructions in the literature that represent partial spreads in a way analogous to our construction with degree $b=2$.

\section{Computational Results on Ranks and EA-Equivalence for $n=8$}
\label{sec:comp-res}

To investigate more in detail the bent functions induced by our LRS construction, we performed a computer search for $n=6$ and $n=8$ variables, with polynomials of degrees $b=1,2$, generating both $\mathcal{PS}^-$ and $\mathcal{PS}^+$ functions. 
Recall that while for $\mathcal{PS}^-$ functions always have degree $n/2$ in general, $\mathcal{PS}^+$ ones can also have different degrees, but this does not seem to be the case with the functions of our LRS construction, judging from our experiments. Indeed, the first interesting remark of our computer search is that \emph{the algebraic degree of the generated functions is always $n/2$}, also in the $\mathcal{PS}^+$ case. It is known that up to $n=6$ variables, all bent functions of degree $n/2$ belong to the completed Maiorana-McFarland class~\cite{polujan20}. Therefore, the smallest interesting case to consider concerning EA-equivalence is $n=8$ variables.

As a first assessment, we generated all $\mathcal{PS}^-$ functions by using families of coprime polynomials of degree $b=1$. Although by Lemma~\ref{lm:psap-irr-1}, we know that all such functions are in $\mathcal{PS}_{ap}$ and coincide with the Desarguesian spread class, we computed their ranks to have independent verification of the count reported by Weng et al.~\cite{weng07}. In this case, we have $m=l=4$ and $t=2^{m-1}=8$. Hence, to construct a function from the Desarguesian spread, we need $8$ coprime polynomials of degree $d=1$ with coefficients over $\F_{2^4}$. Since there are $16$ such polynomials (i.e., $2^{4}-1$ irreducible polynomials with nonzero constant term and the polynomial $X$), one can obtain $\binom{16}{8} = 12870$ $\mathcal{PS}_{ap}$ functions with our construction. Table~\ref{tab:dps-dist} reports the distribution of the 2-ranks for all such functions, and indeed it coincides with the table given by the authors of~\cite{weng07}.
\begin{table}[tb]
	\caption{Distribution of 2-ranks for bent functions of $n=8$ variables in the Desarguesian spread, obtained through the LRS construction with irreducible polynomials of degree $b=1$ over $\F_{2^4}$. The bold value corresponds to the upper bound for the rank of a Maiorana-McFarland function.}
	\centering
	\begin{tabular}{cr}
		\toprule
		Rank & \#Functions \\
		\midrule
		{\bfseries 30}   & 270 \\
		36   & 2160 \\
		40   & 1080 \\
		42   & 9360 \\
		\midrule
		Total & 12870 \\
		\bottomrule
	\end{tabular}
	\label{tab:dps-dist}
\end{table}
The upper bound on the rank of a Maiorana-McFarland function of $n=8$ variables given in~\cite{weng07} is $2^{m+1}-2 = 30$ for $n=8$ variables. Hence, one can see from Table~\ref{tab:dps-dist} that most of the functions in the Desarguesian spread are inequivalent to Maiorana-McFarland functions.

Next, we focused our attention on coprime polynomials of degree $b=2$. As we discussed in Section~\ref{sec:eq-ds}, this case is not directly amenable to the Desarguesian spread, and it is therefore an interesting candidate to find potentially new $\mathcal{PS}^-$ and $\mathcal{PS}^+$ functions. By Theorem~\ref{thm:char-pol}, we have $w=3$, $t=8$, and $l=2$. Consequently, a $\mathcal{PS}^-$ bent function is obtained by finding a set of eight pairwise coprime polynomials over $\F_{4}$ of degree $2$. Let $\F_4 = \{0,1, \alpha, \alpha^2\}$, where $\alpha$ is a root of an irreducible polynomial $p(X) \in \F_2[X]$ of degree $2$. Then, by Gauss's formula, there are six irreducible polynomials of degree $2$ over $\F_4$:
\begin{align*}
p_1(X) &= X^2 + \alpha^2 X + \alpha^2 \enspace ,\\
p_2(X) &= X^2 + \alpha^2 X + 1 \enspace ,\\
p_3(X) &= X^2 + \alpha X + \alpha \enspace ,\\
p_4(X) &= X^2 + X + \alpha^2 \enspace ,\\
p_5(X) &= X^2 + \alpha X + 1 \enspace ,\\
p_6(X) &= X^2 + X + \alpha \enspace .
\end{align*}
These polynomials are, of course, pairwise coprime since they are irreducible. Let us denote them by $\mathcal{I}_2 = \{p_1,p_2,p_3,p_4,p_5,p_6\}$. Further, there are three irreducible polynomials of degree $1$ and nonzero constant term over $\F_4$ that can be squared to obtain polynomials of degree $2$ that are coprime among themselves and with those in $\mathcal{I}_2$:
\begin{align*}
p_7(X) &= (X+1)^2 = X^2 + 1 \enspace ,\\
p_8(X) &= (X+\alpha)^2 = X^2 + \alpha^2 \enspace ,\\
p_9(X) &= (X+\alpha^2)^2 = X^2 + \alpha \enspace .
\end{align*}
Analogously, we denote by $\mathcal{I}_1^2$ the set $\{p_7,p_8,p_9\}$. Further, we can augment our set with the polynomials $1$ and $X^2$. Although the former is not of degree $2$ and the latter does not have a constant term, they are coprime with all polynomials in $\mathcal{I}_1 \cup \mathcal{I}_2^2$.

Finally, we can take the $\binom{3}{2} = 3$ pairs of $\mathcal{I}_1$ and multiply the polynomials in them, obtaining:
\begin{align*}
p_{10}(X) &= (X+1)(X+\alpha^2) = X^2 + \alpha X + \alpha^2 \enspace ,\\
p_{11}(X) &= (X+1)(X+\alpha) = X^2 + \alpha^2 X + \alpha \enspace ,\\
p_{12}(X) &= (X+\alpha)(X+\alpha^2) = X^2 + X + 1 \enspace ,
\end{align*}
with $\mathcal{I}_{1,1} = \{p_{10},p_{11},p_{12}\}$. These three polynomials are not pairwise coprime among themselves, but each of them is relatively prime to all polynomials in $\mathcal{I}_2 \cup \{1,X^2\}$, and to exactly one polynomial in $\mathcal{I}_1^2$. Summarizing, for the $\mathcal{PS}^-$ case, we can construct $174$ functions with the following families of $t=8$ pairwise coprime polynomials:
\begin{itemize}
	\item $\binom{11}{8} = 165$ subsets of $8$ elements in the union $\mathcal{I}_2 \cup \mathcal{I}_1^2 \cup \{1, X^2\}$.
	\item $3$ families obtained by adjoining to $\mathcal{I}_2$ one element from $\mathcal{I}_1^2$ and one from $\mathcal{I}_{1,1}$, so that these last two polynomials are coprime, i.e., $\mathcal{I}_2 \cup \{p_7,p_{12}\}$, $\mathcal{I}_2 \cup \{p_8,p_{10}\}$, and $\mathcal{I}_2 \cup \{p_9,p_{11}\}$.
	\item $6$ families obtained by adding to $\mathcal{I}_2$ one element from $\{1,X^2\}$ and one element from $\mathcal{I}_{1,1}$, i.e., $\mathcal{I}_2 \cup \{1,p_{10}\}$, $\mathcal{I}_2 \cup \{1,p_{11}\}$, $\mathcal{I}_2 \cup \{1,p_{12}\}$, $\mathcal{I}_2 \cup \{X^2,p_{10}\}$, $\mathcal{I}_2 \cup \{X^2,p_{11}\}$, and $\mathcal{I}_2 \cup \{X^2,p_{12}\}$.
\end{itemize}
Similarly, we can obtain $64$ $\mathcal{PS}^+$ functions by the following families of $t=9$ pairwise coprime polynomials:
\begin{itemize}
	\item $\binom{11}{9} = 55$ subsets of $9$ elements in the union $\mathcal{I}_2 \cup \mathcal{I}_1^2 \cup \{1, X^2\}$.
	\item $3$ families obtained by adjoining to $\mathcal{I}_2 \cup \{1,X^2\}$ one element from $\mathcal{I}_{1,1}$, i.e., $\mathcal{I}_2 \cup \{1,X^2\} \cup \{p_{10}\}$, $\mathcal{I}_2 \cup \{1,X^2\} \cup \{p_{11}\}$, and $\mathcal{I}_2 \cup \{1,X^2\} \cup \{p_{12}\}$.
	\item $6$ families obtained by adding to $\mathcal{I}_2$ one element from $\{1,X^2\}$, one from $\mathcal{I}_{1}^2$ and one element from $\mathcal{I}_{1,1}$ so that these last two are coprime. These correspond to the families $\mathcal{I}_2 \cup \{1,p_7,p_{12}\}$, $\mathcal{I}_2 \cup \{1,p_8, p_{10}\}$, $\mathcal{I}_2 \cup \{1,p_9,p_{11}\}$, $\mathcal{I}_2 \cup \{X^2,p_7,p_{12}\}$, $\mathcal{I}_2 \cup \{X^2,p_8, p_{10}\}$, and $\mathcal{I}_2 \cup \{X^2,p_9,p_{11}\}$.
\end{itemize}

Table~\ref{tab:deg2-dist} reports the distribution of the ranks for the $\mathcal{PS}^-$ and $\mathcal{PS}^+$ functions obtained from the families of polynomials described above.

\begin{table}[t]
	\caption{Distribution of 2-ranks for $\mathcal{PS}^-$ and $\mathcal{PS}^-$ bent functions of $n=8$ variables obtained through the LRS construction with coprime polynomials of degree $b=2$ over $\F_{4}$. The bold value corresponds to the upper bound for the rank of the Desarguesian bent function.}
	\centering
	\begin{tabular}{ccr}
		\toprule
		Type & Rank & \#Functions \\
		\midrule
		\multirow{5}{*}{$\mathcal{PS}^-$} & 36 & 20 \\
		& 40 & 24 \\
		& {\bfseries 42} & 10 \\
		& 44 & 60 \\
		& 46 & 60 \\
		& Total & 174 \\
		\midrule
		\multirow{2}{*}{$\mathcal{PS}^+$} & 40 & 45 \\
		& 44 & 19 \\
		& Total & 64 \\
		\bottomrule
	\end{tabular}
	\label{tab:deg2-dist}
\end{table}
\bigskip
The first significant observation that can be drawn from the table is that \emph{none of these bent functions is equivalent to a Maiorana-McFarland function}, since the smallest rank is $36$. 
It is even more interesting to observe that \emph{many functions are inequivalent to the ones induced by the Desarguesian spread}, namely those reaching a rank higher than 42. In particular, our computer search found $60$ $\mathcal{PS}^-$ functions of rank 44 and $60$ of rank 46, and 19 $\mathcal{PS}^+$ functions of rank 44. Thus, we can conclude that our LRS construction can generate bent functions that are not EA-equivalent neither to Maiorana-McFarland functions nor to Desarguesian spread ones. While this is not sufficient to conclude that we found a class of previously unknown bent functions, we consider it as the first step toward that goal. Hopefully, our results will motivate further research in this direction.

\section{Conclusions and Perspectives}
\label{sec:conclusions}

This paper described a method to construct bent functions from linear recurring sequences. The construction leverages on the subspaces spanned by linear mappings defined by a family of LRS. In particular, we proved that if the polynomials defining the linear recurrence equations are pairwise coprime, the kernels of the corresponding linear mappings have a pairwise trivial intersection. This result depends on the observation that the superposition of two LRS mappings is the Sylvester matrix associated with their polynomials, which is invertible if and only if the polynomials are coprime. Consequently, the kernels induced by a family of LRS subspaces whose polynomials are pairwise coprime form a partial spread, and thus a bent function in the class $\mathcal{PS}$.

The key question concerning our LRS construction is to determine when a large enough family of LRS kernels exists, depending on the number of variables of the function, the degree of the polynomials and the extension field of their coefficients. Assuming that all polynomials have a nonzero constant term, we showed that such families exist if and only if the degree of the polynomials is either $1$ or $2$, and we derived the counting formulas for both cases. We then remarked that at least two other polynomials can always be added to these families, namely $X^b$ and $1$. This allows one to obtain also $\mathcal{PS}^+$ functions and, in certain situations, to employ families of polynomials with degrees larger than 2. We then proved that our LRS construction coincides with the Desarguesian partial spread when the degree of the involved polynomials is $b=1$, and thus the functions obtained in this case all belong to the class $\mathcal{PS}_{ap}$. Therefore, candidates for potentially new bent functions generated by our construction should be sought with polynomials of degree $b=2$.

After remarking that the bent functions of $n=6,8$ variables given by our LRS construction always have maximal degree $n/2$ (even for $\mathcal{PS}^+$ ones), we performed a computational analysis of the $2$-ranks of the functions for the $n=8$ case, to determine the number of equivalence classes and their sizes. In particular, for degree $b=1$, we verified the rank distribution reported by Weng et al.~\cite{weng07} for bent functions in the Desarguesian spread, remarking that most of them are not EA-equivalent to any Maiorana-McFarland function. For degree $b=2$, we generated both $\mathcal{PS}^-$ and $\mathcal{PS}^+$ types of functions and remarked that many of them have a rank greater than $42$, which means that they are not EA-equivalent to functions in the Desarguesian spread either. Hence, such bent functions are the most promising candidates to be potentially novel.

There are several open questions to address regarding this LRS construction in future research. The first interesting direction is to investigate more in detail the functions obtained by polynomials of degree $b=2$. Indeed, although we showed that many of them are inequivalent to both Maiorana-McFarland and Desarguesian spread functions, it could still be the case that they are EA-equivalent to some other known classes. To this end, it would be interesting to compare our functions to those generated by other partial spread-based constructions, a list of which can be found in~\cite{mesnager16}. Besides computing the $2$-rank, employing more discriminating invariants would also be interesting. These include, for instance, the \emph{Smith normal form} of the development of the graph $G_f$ of a Boolean function $f$, which is used by Polujan and Pott in~\cite{polujan20} to classify homogeneous cubic bent functions.

We conclude by discussing the connection of our LRS construction with the cellular automata (CA) approach that we adopted in~\cite{gadouleau20}. Our initial idea was to start from a recent construction of \emph{Mutually Orthogonal Latin Squares} (MOLS) based on linear CA that we set forth in~\cite{mariot20}. A cellular automaton can be defined as a shift-invariant vectorial transformation, where the same \emph{local rule} is applied at all sites (or cells) of the input array. If the local rule is linear, then the CA global function is defined by a transition matrix with the same form of the matrix in Equation~\eqref{eq:ca-matr-f}. In particular, the CA global function may be regarded as the linear map induced by an LRS, with Equation~\eqref{eq:rule} representing the application of the local rule on the $i$-th cell of the input.

The authors of~\cite{mariot20} first showed that such a linear CA $F: \F_q^{2(d-1)} \to \F_q^{d-1}$ defines a Latin square of order $q^{d-1}$ if and only if the leftmost and rightmost coefficients $a_0,a_{d-1}$ of its local rule are not null. Further, they proved that the Latin squares generated by two such CA are orthogonal if and only if the polynomials associated with their local rules are relatively prime. Thus, determining the maximum size of a family of pairwise coprime polynomials of degree $d-1$ and the nonzero constant term is equivalent to finding the size of the largest family of MOLS of order $q^{d-1}$ induced by linear CA.

The connection between MOLS generated by linear CA and bent functions traces back to a theorem proved by Bush~\cite{bush73}, where he showed that a large enough orthogonal array (OA, which is equivalent to a set of MOLS) could be used to define a Hadamard matrix. It is well known that a Boolean function is bent if and only if the polar form of its translate design is a Hadamard matrix. What we proved in~\cite{gadouleau20} is that the Hadamard matrix defined by the MOLS of a family of linear CA indeed has the translate design structure required for a bent function. This result is basically the ``CA version'' of Theorem~\ref{thm:ps-lrs} proved in the present manuscript.

The characterization through kernels of LRS is clearly a much more compact way to describe our construction than the CA approach, and it is also more general. Indeed, in this paper, we focused on the assumption that the feedback polynomials of the LRS have a nonzero constant term to leverage on the counting results proved in~\cite{mariot20} for CA-based MOLS. However, Lemma~\ref{lm:trv-int} does not need this hypothesis to characterize LRS kernels with a trivial intersection, which is what matters in the end to construct a partial spread. In particular, one can use \emph{any} family of pairwise coprime polynomials with degree $b$, regardless of their constant term. This is enough to guarantee that the associated Sylvester matrix is invertible. We implicitly dropped this assumption by augmenting our families with the polynomials $X^b$ and $1$ since they are easily seen to be coprime with all other polynomials. However, besides those analyzed here, several other families of coprime polynomials can be considered. We plan to investigate this issue in future research, as we suspect that this would simplify the counting results reported in Section~\ref{sec:count}, by using the $q$-to-1 relationship between non-coprime and coprime pairs of polynomials over $\F_q$ proved in~\cite{benjamin07}.

\subsection*{Acknowledgements}
	The authors wish to thank the anonymous reviewers for their comments to improve the previous version of this paper~\cite{gadouleau20}, as well as pointing out the connection between linear CA and subspaces of linear recurring sequences and the use of rank invariants to classify our bent functions.

\subsection*{Data Availability} The experimental data discussed in this paper (including the truth tables of the functions generated through the LRS construction and their ranks) are available at \url{https://github.com/rymoah/bent-functions-lrs}.

\bibliographystyle{abbrv}      
\bibliography{bibliography}   

\end{document}